\documentclass[11pt]{article}
\pdfoutput=1
\usepackage{jcapmod}
\usepackage{booktabs}
\usepackage[english]{babel}
\usepackage{amsmath,amssymb,amsbsy,amstext, amsthm, simplewick}
\usepackage{hyperref}
\usepackage{graphicx}
\usepackage{amsfonts}
\usepackage{amssymb}
\usepackage[small]{caption}
\usepackage{upgreek}
 \usepackage{exscale,relsize}
\usepackage{soul}
\usepackage{slashed}

\RequirePackage{color}

\newcommand{\beqn}{\begin{equation}}
\newcommand{\eeqn}{\end{equation}}
\newcommand{\req}[1]{Eq.\,\eqref{#1}}
\newcommand{\mpl}{M_{\mathrm{pl}}}

\usepackage{colortbl}
\definecolor{rp}{cmyk}{0.2, 1, 0.6, 0}
\definecolor{green2}{cmyk}{0, 1, 0.5, 0}
\definecolor{lightgreen}{cmyk}{0.2, 0, 0.2, 0.2}
\definecolor{lightgray}{cmyk}{0.1,0.2,0,0.1}
\definecolor{lightgray2}{cmyk}{0.4,0.4,0,0.8}
\definecolor{black}{cmyk}{1.0,1.0,1.0,1.0}

\allowdisplaybreaks[1]

\usepackage{colortbl}
\definecolor{lightgreen}{cmyk}{0.2, 0, 0.2, 0.2}
\definecolor{lightgray}{cmyk}{0.1,0.2,0,0.1}
\definecolor{lightgray2}{cmyk}{0.1,0.1,0,0.1}

\makeatletter
\newlength{\apb@width}
\newcommand{\autoparbox}[2][c]{\settowidth{\apb@width}{#2}\parbox[#1]{\apb@width}{#2}}

\makeatother

\numberwithin{equation}{section}

\def\beq{\begin{equation}}
\def\eeq{\end{equation}}

\def\bea{\begin{eqnarray}}
\def\eea{\end{eqnarray}}

\def\beq{\begin{equation}}
\def\eeq{\end{equation}}
\def\bea{\begin{eqnarray}}
\def\eea{\end{eqnarray}}

\def\0{{\boldsymbol 0}}

\DeclareRobustCommand{\SkipTocEntry}[4]{}

\begin{document}

\begin{titlepage}

\setcounter{page}{1} \baselineskip=15.5pt \thispagestyle{empty}

\bigskip\

\vspace{1cm}
\begin{center}

{\fontsize{20}{28}\selectfont  \sffamily \bfseries Effective Field Theory of Broken Spatial Diffeomorphisms}

\end{center}

\vspace{0.2cm}

\begin{center}
{\fontsize{14}{30}\selectfont   Chunshan Lin}
\end{center}

\begin{center}
\textsl{ Yukawa Institute for Theoretical Physics, Kyoto University \\
Kavli Institute for the Physics and Mathematics of the Universe (WPI), University of Tokyo}
\
\end{center}

\begin{center}
{\fontsize{14}{30}\selectfont   Lance Z. Labun}
\end{center}

\begin{center}
\textsl{ Department of Physics, The University of Texas, Austin, TX 78712, USA\\
Department of Physics, National Taiwan University, Taipei 10617, Taiwan\\
Leung Center for Cosmology and Particle Astrophysics (LeCosPA),
National Taiwan University, Taipei 10617, Taiwan}
\
\end{center}

\vspace{1.2cm}
\hrule \vspace{0.3cm}
\noindent {\sffamily \bfseries Abstract} \\[0.1cm]
We study the low energy effective theory describing gravity with broken spatial diffeomorphism invariance. In the unitary gauge, the Goldstone bosons associated with broken diffeomorphisms are eaten and the graviton becomes a massive spin-2 particle with 5 well-behaved degrees of freedom. In this gauge, the most general theory is built with the lowest dimension operators invariant under only temporal diffeomorphisms.  Imposing the additional shift and SO(3) internal symmetries,  we analyze the perturbations on a FRW background. At linear perturbation level, the observables of this theory are characterized by five parameters, including the usual cosmological parameters and  one additional coupling constant for the symmetry-breaking scalars.  In the de Sitter and Minkowski limit, the three Goldstone bosons are supermassive and can be integrated out, leaving two massive tensor modes as the only propagating degrees of freedom.  We discuss several examples relevant to theories of massive gravity.
\vskip 10pt
\hrule

\vspace{0.6cm}
 \end{titlepage}

 \tableofcontents

\newpage 

\addcontentsline{toc}{section}{References}

\section{Introduction}

In general relativity, space-time diffeomorphism invariance is the local symmetry principle underlying gravitational interactions. One of most profound physical implications is the equivalence  principle \cite{diff1}\cite{diff2}.  However, on a specified space-time background, one or more of the diffeomorphisms are generally broken by gauge fixing, and the pattern of symmetry breaking constrains the low energy degrees of freedom and dynamics on that background.
For instance, our expanding universe can be considered as a temporal diffeomorphism breaking system, because the future always looks different from the past. Theories of gravity with temporal diffeomorphism breaking have been extensively studied in the literature, e.g. k-essence \cite{ArmendarizPicon:1999rj}, the effective field theory of inflation \cite{Cheung:2007st}\cite{Weinberg:2008hq}, ghost condensation \cite{ArkaniHamed:2003uy},  Horava gravity \cite{Horava:2009uw},  generalized Horndeski theories\,\cite{Gleyzes:2014dya}\cite{Gao:2014soa}\cite{Lin:2014jga}\cite{Gao:2014fra} and so on.

Spatial diffeomorphism breaking is also important for the description of our universe: an everyday example is the description of low energy excitations of solids (phonons), which can be derived as the theory of broken spatial diffeomorphism invariance in which the phonons are the Goldstone bosons \cite{Leutwyler:1996er}\cite{Nicolis:2015sra}.  With the addition of a $U(1)$ symmetry to conserve particle number, the theory of broken spatial diffeomorphisms describes ``superfluid solids'' (``supersolids'') \cite{0501658v2}.  In these systems, spatial diffeomorphism invariance is a hidden symmetry that is evidenced by the constrained form of the Goldstone bosons' interactions.  At solar system and cosmological scales, spatial diffeomorphism invariance is a relevant symmetry in that these systems are accurately described by general relativity.  However, the unexplained origins of inflation, the end of inflation and the late time accelerated expansion keep open the possibility that general relativity is modified at the largest and smallest length scales.  
It is therefore interesting to ask how broken spatial diffeomorphisms impact cosmological dynamics.    

In this work, we develop the effective theory for the long-wavelength ($k/a\sim H$ the Hubble constant) degrees of freedom in the presence of broken spatial diffeomorphisms.
As in the condensed matter examples, spatial diffeomorphism invariance can be broken by non-gravitational interactions.  Field theory provides a mechanism in the form of soliton field configurations, such as the hedgehog solution
\begin{eqnarray}\label{hedgehog}
\phi^a=f(r)\frac{x^a}{r}~,~~~~~a=1,2,3,
\end{eqnarray}
which describes a monopole in an $SU(2)$ gauge theory that is spontaneously broken down to $U(1)$.  Here $a$ is the internal index when it is written as the superscript of scalar fields and is the spatial index when the superscript of coordinates. Taking into account gravity, this configuration of the $\phi^a$ fields breaks spatial diffeomorphisms, and in this case, translation and rotation symmetry are also broken to subgroups by fixing a preferred origin of the monopole. This background field configuration has been implemented to produce an inflationary phase in a model known as ``topological inflation'', given that the size of monopole is greater than the Hubble radius in the early universe \cite{Linde:1994hy}\cite{Vilenkin:1994pv}.  The field configuration \req{hedgehog} is not the unique way to break spatial diffeomorphisms, and we will consider a more minimal way that preserves the translation and rotation symmetries.  

The low energy description of broken spatial diffeomorphisms exhibits three Goldstone bosons, scalar fields $\phi^a$ which physically can be thought of as measuring spatial position.  In unitary gauge, these ``ruler fields" are identified with the coordinates,
\beqn\label{spatialcond}
\phi^a=x^a, \qquad  a=1,2,3.
\eeqn
Translation and rotation invariance are preserved by implementing a shift symmetry $\phi^a\to\phi^a+c^a$ for constants $c^a$ and an $SO(3)$ internal symmetry in the triplet $\phi^a$. The scalars $\phi^a$ select a frame of reference, a background against which to measure perturbations.  To restore the Goldstone bosons as dynamical degrees of freedom, we add a fluctuating component to the field 
\beqn\label{condpis}
\phi^a\to x^a+\pi^a
\eeqn
with $\pi^a$ transforming under spatial diffeomorphisms opposite to the spatial coordinates and thus furnishing a nonlinear realization of the symmetry (known as the St\"uckelberg trick).  To see how this describes a solid, think of the scalar functions $\phi^a(x)$ as locating each volume element or lattice site in space.  In the long-wavelength limit $\lambda\gg$ lattice spacing, inhomogeneity at the sites is smoothed over, and fluctuations of the $\phi^a$ correspond to fluctuations of the site locations, i.e. phonons \cite{Leutwyler:1996er,Nicolis:2015sra}. 

Broken spatial diffeomorphism invariance is interesting in the context of gravitational theory, because it generates a mass for the graviton.  This is easy to understand seeing that the presence of a fixed frame (one may think of a lattice) admits the propagation of additional compressional and rotational modes, which are the longitudinal modes of the graviton.  
The structure of the broken spatial  diffeomorphism theory thus helps understand how to construct a general self-consistent theory of massive gravity. 
Indeed, it is a basic question in classical field theory whether an analog of Higgs mechanism exists that can give gravitons a small but non-vanishing mass.  Experimentally, we do not know how gravity behaves at distances longer than $\sim 1$\,Gpc, and the extremely tiny energy-scale associated with the cosmic acceleration\cite{981,982} hints that gravity might need to be modified at such large scale.

The theoretical and observational consistency of massive gravity has been a longstanding problem.  In the pioneering attempt in 1939 by Fierz and Pauli \cite{Fierz1939}, the simplest extension of GR with a linear mass term suffers from  the van Dam-Veltman-Zakharov discontinuity \cite{vdvz1}\cite{vdvz2}, giving rise to different predictions for the classical tests in the vanishing mass limit.  This problem can be alleviated by introducing nonlinear terms \cite{Vainshtein}. However, in 1972,  Boulware and Deser pointed out that a ghost generally reappears at the nonlinear level, which spoils the stability of the theory \cite{bdghost}. 
Inspired by effective field theory in the decoupling limit \cite{ArkaniHamed:2002sp}, people have learned that in principle the Boulware-Deser ghost can be eliminated by construction \cite{deRham:2010ik}\cite{deRham:2010kj}. This theory is now dubbed dRGT gravity. When we apply dRGT gravity to cosmology, a self-accelerating solution is found for the open FRW universe \cite{Gumrukcuoglu:2011ew}.  However, follow-up cosmological perturbation analysis found a new ghost instability among the 5 gravitational degrees of freedom \cite{Gumrukcuoglu:2011zh}\cite{DeFelice:2012mx}\cite{DeFelice:2013awa}\cite{DeFelice:2013bxa}\cite{Khosravi:2013axa}, and further dRGT gravity might suffer from acausality problems \cite{Deser:2012qx}\cite{Deser:2013eua}.  The dRGT ghost instability can be eliminated at the expense of introducing a new degree of freedom \cite{DeFelice:2013tsa}\cite{DeFelice:2013dua}.  In this context, it is interesting to search for a simpler and self-consistent massive gravity theory, as an alternative to the Fierz-Pauli family of theories. 

An alternative way to realize a massive gravity theory is to break the Lorentz symmetry of vacuum configuration, in addition to the space-time diffeomorphisms.
A broad class of Lorentz-symmetry breaking massive gravity theories have been discussed in Refs.  \cite{Dubovsky:2004sg}\cite{Rubakov:2004eb}\cite{Bluhm:2007bd}\cite{Comelli:2013txa}\cite{Blas:2014ira}.  Among these theories, a simple example is the spatial condensation scenario \req{spatialcond}; the non-vanishing spatial gradient breaks 3 spatial diffeomorphisms, while temporal diffeomorphism, translational and rotational invariance are preserved \cite{Lin:2013aha}\cite{Lin:2013sja}. Previous analyses focused on linear theory in the decoupling limit.  As we will see below, the theory becomes degenerate in the Minkowski space time.  On FRW backgrounds, there are exactly 5 degrees of freedom in the theory. In the unitary gauge, the graviton eats the Goldstone excitations $\pi^a$ in \req{condpis} and becomes a massive spin-2 particle, with 5 massive modes in the spectrum.  

The resulting theory has several interesting applications. For instance, the graviton mass removes the IR divergence in graviton scattering \cite{Lin:2013sja}, and leaves an interesting imprint on CMB primordial tensor spectrum \cite{Cannone:2014uqa}.  A viable massive gravity theory also provides the basis for holographic study of dissipative systems\,\cite{Davison:2013jba,Blake:2013bqa,Blake:2013owa}.  Several other gravitational phenomena associated with broken spatial diffeomorphisms have also been discussed in the literature \cite{Dubovsky:2005xd}\cite{Endlich:2010hf}\cite{Endlich:2012pz}\cite{Ballesteros:2012kv}\cite{Bartolo:2013msa}\cite{Ballesteros:2013nwa}\cite{Akhshik:2014gja}\cite{Kouwn:2014aia}\cite{Pearson:2014iaa}, without relating them explicitly to the massive gravity aspect of the theory.  For example, by tuning the form of higher order interactions, Ref. \cite{Endlich:2012pz} builds a model of inflation, calling it ``solid inflation'', in which they calculate the two- and three-point correlation of primordial perturbation.  Our analysis helps understand why the sound speeds of scalar and vector modes are related in such a theory.

In this paper, we study the general gravitational action for broken spatial diffeomorphisms by constructing the appropriate low energy effective field theory. The effective field theory approach describes a system through the lowest dimension operators compatible with the underlying symmetries. Usually, when we study a gravitational system, we first write down a general covariant action, and space-time diffeomorphisms are broken ``spontaneously'' after solving the equation of motion. However, in this paper to learn more of the structure of the theory and resulting character of massive gravity, we start by writing down the most general gravitational action compatible with spatial diffeomorphisms breaking in the unitary gauge.  We then recover general covariance by performing a change of spatial coordinates $x^i\to x^i+\xi^i$ and promote the parameters $\xi^i$ to Goldstone bosons which transform opposite to the spatial coordinates $\pi^a\to\pi^a-\xi^a$ under spatial diffs.

This paper is organised as follows:  In section \ref{sec:ugaugeaction}, 
before constructing the specific effective theory, we discuss the general set of terms allowable in unitary gauge. Because the unitary gauge action, in its initial background-independent form, does not make explicit the dynamical degrees of freedom, we must carefully select the terms so as to preserve the 5 desired degrees.
Then in section \ref{sec:FRWaction}, we specify to the FRW background, discuss the physical scales of interest, including different requirements during inflation and late-time and requirements for the perturbativity of the theory.  Restricting to $SO(3)$ rotational symmetry and shift symmetry, we determine the effective action in the FRW universe, and analyse all scalar, vector, and tensor degrees of freedom.  In section V, we present several examples of the applications of our formalism. Conclusion and discussion will be in the final section VI.  In this paper we use the $(-,+,+,+)$ convention in the space time metric.

\section{Generic action in unitary gauge}\label{sec:ugaugeaction}

To help show how a theory of broken spatial diffeomorphisms is a theory of massive gravity, we first discuss constructing the Lagrangian in the unitary gauge, in which we only have metric degrees of freedom.  When we analyze perturbations, we identify which metric components become dynamical, corresponding to the longitudinal polarizations of the graviton, and thus in a given allowed operator we can track the real degrees of freedom. This is important because much previous study of modified gravity theories has shown that many forms of higher derivative operators lead to new degrees of freedom.  To preserve exactly 5 dynamical degrees of freedom (2 graviton polarizations + 3 goldstone bosons), the set of operators must be additionally constrained.  As these constraints apply to the construction in any background, it is worth investigating allowable terms in a ``generic'' action in unitary gauge, before specifying the background solution (and with it the relevant scales).  This facilitates construction in other backgrounds.

In unitary gauge, the action is only invariant under the temporal diffeomorphisms. There is a preferred spatial frame generated by the space like gradient of scalar fields, $g^{\mu\nu}\partial_{\mu}\phi^a\partial_{\nu}\phi^bh_{ab}>0$, where $\phi^a(t,\textbf{x})$ generally  is a function of space and time, and $h_{ab}$ is the internal metric of scalar fields' configuration. In the unitary gauge, the spatial frame $x^a$ is chosen to coincide with $\phi^a$, 
\begin{eqnarray}\label{scond}
\langle\phi^a\rangle=x^a,~~~~~~~~a=1,2,3.
\end{eqnarray}
They transform as the scalars under the residual diffeomorphisms, so that the additional degrees of freedom are in the space-time metric. 

Going systematically through the geometric objects, we have:
\begin{enumerate}
\item Terms that are invariant under all diffeomorphisms.  These include polynomials of the Riemann tensor $R_{\mu\nu\kappa\lambda}$ and its covariant derivatives contracted to give a scalar. However, many such terms introduce additional unwanted degrees of freedom and/or break temporal diffeomorphisms. For instance, by doing a conformal transformation, $\mathcal{R}^2$ is equivalent to Einstein gravity plus a scalar field with non-trivial potential. To remain within the effective theory and its degrees of freedom, we need only the linear term in Ricci scalar $\mathcal{R}$ to the order considered.  
\item Any scalar function of coordinates $x^a$, as well as their covariant derivatives. In the unitary gauge, $\nabla_{\mu}x^a=\delta_{\mu}^a$.  Higher than second order derivatives generally give rise to extra modes and the classical Ostrogradski ghost instability. However, provided we have a viable perturbative expansion, in which higher dimensioned operators including higher derivative operators are supporessed by a high scale $\Lambda$, the typical mass of the Ostrogradski ghost is at or above the cut-off scale of our effective field theory. For this reason, the would-be ghosty modes are non-dynamical and can be integrated out at the low energy scale \cite{Weinberg:2008hq}. Consider for instance the scalar field theory with higher order derivative terms
\begin{eqnarray}
\mathcal{L}=-\frac{1}{2}\left[\partial_{\mu}\phi\partial^{\mu}\phi+\Lambda^{-2}\left(\square\phi\right)^2\right],
\end{eqnarray} 
for which the propagator reads
\begin{eqnarray}
\Delta(k)=\frac{1}{k^2+\frac{k^4}{\Lambda^2}}=\frac{1}{k^2}-\frac{1}{k^2+\Lambda^2},
\end{eqnarray}
with two propagating degrees. The second appears to have the wrong sign propagator, which is the possible ghosty mode. However, the pole is at $k^2=-\Lambda^2$, which is around the cutoff scale of our theory.  For momenta in the domain of the effective theory $k\ll \Lambda$, this degree of freedom is supermassive and can be integrated out.

\item We can leave free the upper indices $i$ in every tensor. For instance we can use $g^{ij}$, $\mathcal{R}^{ij}$ and $\mathcal{R}^{ijkl}$. However, we notice that $\mathcal{R}^{ij}$ can be rewritten into higher order covariant derivatives of $x^a$ by  partial integration,
\begin{eqnarray}
\mathcal{R}^{aa}=\mathcal{R}^{\mu\nu}\partial_{\mu}x^a\partial_{\nu}x^a=\left(\square x^a\right)^2-\left(\nabla_{\mu}\nabla_{\nu}x^a\right)\left(\nabla^{\mu}\nabla^{\nu}x^a\right)+\text{total derivatives},
\end{eqnarray}
so is $\mathcal{R}^{ijkl}$ term.  Thus $\mathcal{R}^{ij}$ and $\mathcal{R}^{ijkl}$ terms  are irrelevant at low energy scale.
\end{enumerate}
We conclude that the most generic Lagrangian in the unitary gauge is given by 
\begin{eqnarray}
S=\int d^4x\sqrt{-g}F\left(\mathcal{R}, x^a, \nabla_{\mu}, g^{ij}, \mathcal{R}^{ij}, \mathcal{R}^{ijkl}\right),
\end{eqnarray}
where all the free indices inside of the function $F$ must be $i$'s.  The construction so far is general to the extent of a gravitational action respecting time diffeomorphism invariance and excluding degrees of freedom additional to the broken spatial diffeomorphisms.  Additional symmetries must be respected when the theory is considered on a particular background, and in the next section we will discuss the restriction.

\section{Expanding around a FRW background}\label{sec:FRWaction}

\subsection{Scales, power counting and constraints}\label{ssec:pc}

In the present work, we are primarily interested in cosmological phenomenology of the broken spatial diff theory, and hence we are looking at wavelengths of the Hubble scale.  To be consistent with the high degree of spatial isotropy and homogeneity at cosmological lengths, the effective scalars $\pi^a$ have an $SO(3)$ internal symmetry and a shift symmetry $\pi^a\to\pi^a+c^a$.  The shift symmetry requires the Goldstones have only derivative couplings, and consequently the effective theory is essentially an expansion in $k/M$ where $M$ is a scale appearing in association with the higher dimensioned operators.  For the theory to be perturbative in that lower order (lower derivative) operators are more relevant than higher orders, we should have $H\ll \Lambda$ the breakdown momentum scale, which is approximately the momentum scale at which $\pi\pi$ scattering violates unitarity.  $\Lambda$ is related to the $M$s appearing in various higher derivative operators and thus the condition $H\ll\Lambda$ yields constraints on the sizes of the higher order operators.  Conversely, we constrain $M$ considering (lack of) evidence for higher order operators. Satisfying these constraints, power counting derivatives $k/M$ yields a consistent theory because the breakdown scale is parametrically above the dynamical range of the theory.

Note that the nontrivial background, the FRW spacetime, introduces an important scale in the dynamics.    $H$ and $\dot H$ enter by determining the characteristic scale of the background $\phi^a$ fields, and as a consequence their fluctuation components $\pi^a$. As we will show in the subsection \ref{Goldstone}, in the $\dot H\to 0$ limit, the theory becomes strongly coupled and encounters some of the issues well-known to massive gravity on Minkowski backgrounds, perhaps pointing to a deeper physics reason for these issues.  On the other hand, it has been previously proven that the present theory is continuous with GR in the limit $M\to 0$, with the Goldstones decoupling, becoming nondynamical and restoring the full diffeomorphism symmetry \cite{Lin:2013sja}.

Considering the early universe, the kinetic energy of a scalar field with $k\sim H$ is $\sim\sqrt{2\epsilon}\mpl H$ where $\epsilon=-\dot H/H^2$ is the slow roll parameter, meaning its change in amplitude over a Hubble time $H^{-1}$ is $\sim\sqrt{2\epsilon}\mpl$.  Since the triplet of dynamical scalars $\pi^a$ can at most be responsible for inflation, the mass scale suppressing higher order terms must be larger $M\gtrsim\sqrt{2\epsilon}\mpl$ \cite{Weinberg:2008hq}.  As we will show, achieving a self-accelerating cosmology with a minimal form of this theory requires fine-tuning the parameters in such a way that the theory becomes strongly-coupled and loses its meaning.  We can consider instead inflation by independent dynamics, with small $\pi^a$ fluctuations on top.  In this case, the scale of the $\pi^a$ must be smaller than the inflation energy scale, but still large enough to validate an expansion in $k/M$; this scenario is quite natural in the context of GUT-scale inflation and the possibility of topological defects arising from breaking the GUT symmetry group.  In fact, the energy density of the scalar kinetic term dilutes as $g^{\mu\nu}\partial_\mu\pi^a\partial_{\nu}\pi^a\sim a^{-2}$ the same as a network of topological defects \cite{Brandenberger:1993by}\cite{Bucher:1998mh}, making such an identification tempting.  Since the $\pi^a$ are small perturbations to dominant inflationary background, evidence of this scenario must be sought in the gravitational wave signal from inflation, via the (effective) mass for gravitational waves\:\cite{Cannone:2014uqa}.

At late times $z\lesssim$ a few, $H$ is much smaller, and the mass scale of the spatial diff breaking mechanism  certain to be higher, in particular if it is related to any standard model particle or astrophysics.  Considering the present universe, we can constrain the scales via the resulting mass and sound speed.  For non-vanishing graviton mass, the orbits of binary system will decay at a  slightly faster rate than predicted by GR, due to the additional energy loss from in the emission of massive  gravitational waves. The decay rate difference can be roughly estimated as 
$\delta\sim\left(\frac{m_g}{\omega}\right)^2,$
where $\omega$ is the frequency of gravitational waves, which is identical to the inverse of orbital period of binary, and $m_g$ is the graviton mass nowadays. Consider the Hulse-Taylor binary pulsar, PSR B 1913+16, for which the observed orbital decay attributed to gravitational wave emission agrees with the predictions of GR to $0.3\%$,  $\delta<0.3\%$ requires that $m_g<10^{-20}eV\sim 10^{12}H_0$ \cite{Sutton:2001yj}. Non-vanishing graviton mass also changes the propagating speed of gravitational waves, which leads to the upper bound on the graviton mass of $m_g<10^{-23}eV$\cite{Will:2005va}. The detection of gravitational waves in advanced LIGO could then bound the graviton mass potentially all the way down to $m_g<10^{-29}eV$ \cite{Will:2005va}\cite{Will:1997bb}\cite{Berti:2004bd}. See \cite{deRham:2014zqa} for a recent review on the theoretical and experimental aspects of massive gravity.

\subsection{Construction of the Action}
For a given FRW, the enhanced symmetry reduces the set of allowed operators from those discussed in the previous secion. In addition to preserving the unbroken temporal diffeomorphism invariance, we must preserve homogeneity (translation invariance) and isotropy.  
Terms like $x^a$ without covariant derivative operators acting on them are not allowed, because they break homogeneity and isotropy. Instead, $x^a$ terms must appear in the action with derivatives, 
\begin{eqnarray}\label{fidtensor}
f_{\mu\nu}\equiv\partial_{\mu}x^a\partial_{\nu}x^b\delta_{ab},
\end{eqnarray}
and any pair of $\partial_{\mu}x^a$ should be contracted with internal metric $\delta_{ab}$ to maintain the $SO(3)$ spatial rotation symmetry. Each $\partial_{\mu}x^a\partial_{\nu}x^b\delta_{ab}$ is contracted with the metric $g^{\mu\nu}$ and thus gives rise to terms proportional to the trace of spatial metric $g^{ii}$  and cross terms like $g^{ij}g^{ij}$  in the action. At linear perturbation level, the cross terms decompose into the trace sector, which is a function of $g^{ii}$, plus the traceless sector which only appears at quadratic order in the action for perturbations.

As for higher derivative terms, they are ``less relevant" at low energy scale, as we discussed in the section \ref{sec:ugaugeaction}. Thus we only focus on lowest dimensional operators for the time being. The higher order derivative terms will be informatively discussed in section \ref{sechd}.

We can now write down our most generic action in the unitary gauge as follows, 
\begin{eqnarray}\label{gact}
S=\int d^4x\sqrt{-g}\left[\frac{1}{2}M_p^2\mathcal{R}+\Lambda+c\sum_ig^{ii}+...\right],
\end{eqnarray}
where the dots stand for terms which are of at least quadratic order in the fluctuations. The terms $g^{ii}$ is responsible for the ``spatial diffeomorphisms breaking". Notice that the above three terms contain linear perturbations around FRW background. Therefore, the coefficients $\Lambda$ and $c$  will be fixed by requiring that all tadpole terms cancel around the given FRW solution. The differences between models will be encoded into higher order terms. 

To fix the coefficients  $\Lambda$ and $c$ of linear terms, we insert the FRW ansatz for the background space time metric, 
\begin{eqnarray}
ds^2=-dt^2+a^2d\textbf{x}^2.
\end{eqnarray}
The Einstein equations read
\begin{eqnarray}
3M_p^2H^2&=&-3\frac{c}{a^{2}}-\Lambda,\nonumber\\
M_p^2\dot{H}&=&\frac{c}{a^{2}}.
\end{eqnarray}
Solving these two equations for $c$ and $\Lambda$, we have
\begin{eqnarray}
\Lambda&=&-3M_p^2\left(H^2+\dot{H}\right),\nonumber\\
c&=&M_p^2a^2\dot{H}.
\end{eqnarray}
At the first glance, one may worry that temporal diffeomorphism invariance is broken since $c$ and $\Lambda$ are time dependent. However, the time dependence of the coefficients is not sufficient to break temporal diffs. One way to check is to perform a general coordinate transformation $x^{\mu}\to x^{\mu}+\xi^{\mu}$, and then promote all four parameters $\xi^{\mu}$ into  fields, $\xi^\mu\to\pi^\mu(x)$. These four $\pi^\mu$ are the would-be Goldstone bosons associated with broken diffeomorphisms. It is easy to check that there is no dynamical Goldstone boson associated with temporal diffeomorphism breaking, or in other words, temporal diffeomorphisms $t\to t+\xi^0$ remain gauge redundant.

We turn now to the fluctuation operator. The simplest form of operator that starts linear order in fluctuations is
\begin{eqnarray}\label{deltagij}
\bar{\delta} g^{ij}\equiv g^{ij}-3\frac{\sum_k g^{ik}g^{kj}}{\sum_k g^{kk}}.
\end{eqnarray}
To distinguish it from the metric fluctuation $\delta g^{\mu\nu}$, we put a bar over $\delta$.  The trace vanishes up to linear order in fluctuations, $\sum_i\bar\delta_1 g^{ii}=0$, where the subscript 1 on $\bar\delta$ denotes linear order in the expansion. 

The term quadratic in fluctuations can be obtained equivalently as the second order of the operator $\sum_i\bar{\delta}_2 g^{ii}$ or the operator $\sum_{ij}\bar{\delta} g^{ij}\bar{\delta} g^{ij}$.  The two differ only by a factor $a$ which can be absorbed by redefinition of the coefficient, and we do not need to write the $\sum_i\bar{\delta}_2 g^{ii}$ term.

The construction of the operator \req{deltagij} is not unique. For instance, we can equally write
\begin{eqnarray}
&&g^{ij}-3\frac{\sum_{k,l} g^{ik}g^{kl}g^{lj}}{\sum_{k,l} g^{kl}g^{kl}}~,~~~g^{ij}-3\frac{\sum_{k,l,m} g^{ik}g^{kl}g^{lm}g^{mj}}{\sum_{k,l,m} g^{kl}g^{lm}g^{mk}}~,\nonumber\\
&&3\sum_{i,j,k}g^{ij}g^{jk}g^{ki}+2\sum_{i,j,k}g^{ii}g^{jk}g^{jk}-\left(\sum_ig^{ii}\right)^3,~~...
\end{eqnarray}
All of these operators as well as the products among them,  give rise to exactly the same quadratic action as $\bar{\delta} g^{ij}\bar{\delta} g^{ij}$, up to a prefactor of scale factor $a$, which can be absorbed into the coefficient. The reason is quite simple. Suppose we have a general operator $T^{ij}$  constructed out of only the spatial metric $g^{ij}$. Expand the spatial metric around the FRW background and then decompose the linear perturbations into trace part and traceless part, 
\begin{eqnarray}\label{metricdecom}
g^{ij}\equiv a^{-2}\delta^{ij}+\frac{1}{3}\delta g^{kk}\delta^{ij}+\bar{\delta}g^{ij},
\end{eqnarray}
where $\bar{\delta}g^{ij}$ is traceless.  The trace part of the perturbation can be absorbed by redefinition of the scale factor $a$.  Therefore, if we demand that the background part of operator $T^{ij}$ vanishes, the trace part of the perturbation must also vanish, and we have $T^{ij}=f(a)\bar{\delta}g^{ij}$ at linear level, where $f(a)$ is a generic function of scale factor $a$. The quadratic order operator can be constructed out of the product of two $T^{ij}$s, i.e. $\bar{\delta} g^{ij}\bar{\delta} g^{ij}$. On the other hand, note that the trace $T\equiv \sum_iT^{ii}$ vanishes at linear perturbation level and at non-linear level $T\sim \bar{\delta}g^{ij}\bar{\delta}g^{ij}$ again. 
We have thus proven that $\bar{\delta} g^{ij}\bar{\delta} g^{ij}$  is the only quadratic order operator needed.

Putting these elements together, we write the action Eq.(\ref{gact}) as 
\begin{eqnarray}\label{ggact}
S=\int d^4x\sqrt{-g}\left[\frac{1}{2}M_p^2\mathcal{R}-3M_p^2\left(H^2+\dot{H}\right)+M_p^2a^2\dot{H}g^{ii}-M_p^2M_2^2\bar{\delta} g^{ij}\bar{\delta} g^{ij}+...\right],
\end{eqnarray}
where the dots stand for the operators starting from cubic order in fluctuations. For the simplicity of notation, we drop all summation symbols, all repeated indices should be summed up by default.  
Noted that $M_2^2$  could be a generic functions of $g^{ii}$ and thus time dependent.  

\subsection{Action for the Goldstone Bosons}\label{Goldstone}

To exhibit the three Goldstone bosons that nonlinearly restore general covariance, we can perform a broken spatial diffeomorphism  (so called $St\ddot{u}ckelberg$ trick). These three Goldstone bosons are decomposed into 1 scalar mode and 2 vector modes, in addition to the 2 tensor modes in GR. In the unitary gauge, the graviton ``eats'' the three Goldstone bosons and becomes a massive spin-2 particle with 5 polarizations: one helicity-0 mode and two helicity-1 modes in addition to the helicity-2 tensor modes. 

Under the broken spatial diffeomorphism, $x^i\to\tilde{x}^i=x^i+\xi^i$ and $t\to\tilde{t}=t$, the metric $g^{ij}$ transforms as 
\begin{eqnarray}\label{gijdiff}
g^{ij}(x)\to\tilde{g}^{ij}\left(\tilde{x}(x)\right)=\frac{\partial\tilde{x}^i(x)}{\partial x^{\mu}}\frac{\partial\tilde{x}^j(x)}{\partial x^{\nu}}g^{\mu\nu}(x),
\end{eqnarray}
and $d^4x\sqrt{-g}$ is invariant under all space time diffeomorphisms. Recalling the construction of the action in the unitary gauge, $\Lambda$ and $c$ in eq. (\ref{gact}) are functions of $g^{ii}$, and their values are fixed by the background Einstein equations to cancel out tadpole terms. This requires that under the broken diffeomorphisms, $\Lambda$ transforms as 
\begin{eqnarray}
\Lambda\to\Lambda+\frac{d\Lambda}{dt}\left(\frac{dg^{ii}}{dt}\right)^{-1}\delta_{\xi}g+\frac{1}{2!}\left(\frac{dg^{jj}}{dt}\right)^{-1}\frac{d}{dt}\left[\frac{d\Lambda}{dt}\left(\frac{dg^{ii}}{dt}\right)^{-1}\right]\delta_{\xi}g\delta_{\xi}g,
\end{eqnarray}
where $\delta_{\xi}g$ is the variation of $g^{ii}(x)$ under the broken diffeomorphisms, i.e. the trace of eq. (\ref{gijdiff}). Noted we have applied chain rule in the Taylor expansion. The coefficient $c$ transforms in the same way as $\Lambda$.
After the spatial diffeomorphism transformation \req{gijdiff}, the action Eq. (\ref{ggact}) reads
\begin{eqnarray}
S=M_p^2\int d^4x\sqrt{-g}\left[\frac{1}{2}\mathcal{R}-3\left(H^2+\dot{H}\right)+a^2\dot{H}g^{\mu\nu}\frac{\partial\left(x^i+\xi^i\right)}{\partial x^\mu}\frac{\partial\left(x^i+\xi^i\right)}{\partial x^\nu}-\frac{1}{3}\left(2\dot{H}+\frac{\ddot{H}}{H}\right)\left(\partial_i\xi^i\right)^2+...\right].\nonumber\\
\end{eqnarray}
Then we promote the nonlinear parameters $\xi^i$ into scalar fields, $\xi^i\to\pi^a\delta_a^i$, and assign to $\pi^a$ the transformation rule
\begin{eqnarray}\label{pidif}
\pi^a(x)\to\tilde{\pi}^a\left(\tilde{x}(x)\right)=\pi^a(x)-\delta_i^a\xi^i(x).
\end{eqnarray} 
With this definition, the Goldstone scalars non-linearly recover general covariance and describe the fluctuation around the FRW background.

The resulting action for the Goldstones is
\begin{eqnarray}
S=M_p^2\int d^4x\sqrt{-g}&&\left[\frac{1}{2}\mathcal{R}-3\left(H^2+\dot{H}\right)+a^2\dot{H}\left(g^{ii}+2\delta g^{a\mu}\partial_{\mu}\pi^a+g^{\mu\nu}\partial_{\mu}\pi^a\partial_{\nu}\pi^a\right)-\frac{1}{3}\left(2\dot{H}+\frac{\ddot{H}}{H}\right)\left(\partial_a\pi^a\right)^2\right.
\nonumber\\ \label{GSaction}
&&\left.-2M_2^2\partial_{a}\pi^b\bar{\delta} g^{ab}-\frac{2M_2^2}{a^4}\cdot\left(\partial_i\pi^a\partial_i\pi^a+\frac{1}{3}\left(\partial_a\pi^a\right)^2\right)+...\right],
\end{eqnarray}
One should keep in mind that the $\pi$s are the physical excitations over the 3 scalar fields' vevs in Eq.\,(\ref{scond}). Note that in the increased symmetry of the de Sitter limit $\dot H\to 0$, the kinetic term $(\dot\pi^a)^2$ vanishes and the expansion breaks down.

When the energy scale is much greater than the mass of gauge boson, the mixing between longitudinal and transverse components of the gauge field becomes irrelevant (helicity is approximately conserved).  The two sectors decouple and analysis is greatly simplified in the Goldstone language, where there are only interacting scalars. 
As seen in \req{GSaction}, the leading order mixing is determined by $\dot{H}$ or $M_2^2$ and this simplification is also achieved when the wavelength is small enough.  The leading order mixing terms between Goldstone bosons and metric perturbations are 
\begin{eqnarray}
M_p^2\dot{H}\delta g^{aj}\partial_{j}\pi^a~,~~~\text{and}~~~M_p^2M_2^2\partial_{a}\pi^b\bar{\delta} g^{ab}.
\end{eqnarray}
We canonically normalize the action for Goldstone pions and metric perturbations as  $\pi_c^a\sim M_pa\dot{H}^{1/2}\pi^a$ and $\delta g_c^{ij}\sim M_p\delta g^{ij}$, after which the mixing terms read
\begin{eqnarray}\label{mixing}
\dot{H}^{1/2}\delta g_c^{aj}\partial_j\pi_c^a~,~~~\text{and}~~~a^{-1}M_2^2\dot{H}^{-1/2}\partial_{a}\pi_c^b\bar{\delta} g_c^{ab}.
\end{eqnarray}
Note that it is appropriate to compare 3-momentum scales here due to violation of Lorentz invariance. We can see that mixing can be neglected for energies above both of $\dot{H}^{1/2}$ and $M_2^2\dot{H}^{-1/2}$.  As expected, the mixing scale is essentially the mass of dynamical modes on the metric perturbations derived below, as the co-factor $\epsilon^{1/2}$ is absorbed in canonical normalization. After neglecting the mixing, the action of Goldstone Bosons dramatically simplifies to 
\begin{eqnarray}\label{decoupling}
S=M_p^2\int d^4x\sqrt{-g}\left[a^2\dot{H}g^{\mu\nu}\partial_{\mu}\pi^a\partial_{\nu}\pi^a-\frac{1}{3}\left(2\dot{H}+\frac{\ddot{H}}{H}\right)\left(\partial_a\pi^a\right)^2-\frac{2M_2^2}{a^4}\cdot\left(\partial_i\pi^a\partial_i\pi^a+\frac{1}{3}\left(\partial_a\pi^a\right)^2\right)+...\right].\nonumber\\
\end{eqnarray}

Away from the short wavelength approximation, for $\dot{H}/H^2\sim \mathcal{O}(1)$ (also far away from de Sitter), the first coupling term in Eq.\,(\ref{mixing}) is important, and to understand dynamics of modes at the momentum scale of interest $k^2\sim a^2\dot{H}$, a full perturbations analysis is necessary, which will be presented in the next subsection.  In the de Sitter limit $\dot{H}\to0$, the second mixing term diverges, another manifestation of the strong-coupling problem.  We shall see below in Sec.\ref{sec:dSMink} that strong-coupling is avoided with the inclusion of higher derivative terms. If the coefficient $M_2$ is independent of or only weakly dependent on the scale factor, the mixing can become negligible at late time.

Before moving to the perturbations analysis, we take advantage of the ease with which the breakdown scale can be estimated in the Goldstone language.  Since Lorentz invariance is broken, $\Lambda$ can be written as a cutoff in 3-momentum or in energy, which are related approximately by a factor of the sound speed at large $k$.  Reading from \req{GSaction}, the interaction terms should be subdominant compared to the kinetic energy for the theory to be perturbative.  The most stringent constraint on the breakdown scale comes from looking at the 3-$\pi$ coupling inside the $M_2^2\bar\delta g^{ij}\bar\delta g^{ij}$ term.  The $1\to 2$ amplitude becomes of order 1 at the 3-momentum scale 
\beqn\label{breakdown}
\Lambda \sim a^{7/2}\dot H^{3/4}\frac{M_p^{1/2}}{M_2}=a^{7/2}\epsilon^{3/4}H\frac{(HM_p)^{1/2}}{M_2}.
\eeqn
where $\epsilon=-\dot H/H^2$ is the usual slow roll parameter.  Provided $M_2$ is not too close to $\mpl$, this is parametrically higher than the mixing scales given in \req{mixing}.  On the other hand, for the effective theory to be useful in the cosmological context, the breakdown scale should be (much) larger than $H$ so that it effectively describes long wavelength near-horizon dynamics.  The requirement $\Lambda\gg H$ is equivalent to
\beqn\label{Mconstraint}
M_2\ll a^{7/2}\epsilon^{3/4}(HM_p)^{1/2}.
\eeqn
Note that other operators, for example $\bar\delta g^{ij}\bar\delta g^{jk}\bar\delta g^{ki}$ at third order, also contribute to the 3-$\pi$ coupling.  A priori, the corresponding coefficient $M_3^2$ (in obvious notation) is the same order of magnitude as $M_2^2$.  However, when expanding the operator to obtain the $n$-$\pi$ interaction terms, each power of $\bar\delta g$ brings with it a factor $a^{-2}$.  Consequently, the effect of operators that appear at higher order in unitary gauge is typically diluted faster in the expansion of the universe. Only if the associated coefficients scale with a compensating power of $a$ can these higher orders be relevant or worse lead to strong coupling at late time.  As we do not observe a phase transition in the gravitational dynamics (except possibly the end of inflation), we may exclude this possibility and consider that terms higher than the $M_2^2$ term are suppressed.

\subsection{Full Perturbations Analysis}

We have learned that away from short wave length limit, generally the coupling between Goldstone bosons and gravity cannot be omitted. To perform the full perturbations analysis, we first decompose the modes according to helicity and then identify and integrate out non-dynamical metric degrees of freedom. This procedure results in the effective actions for the 5 dynamical modes, though the calculations are somewhat tedious.  

Due to the $SO(3)$ rotational symmetry of our background space time, we can decompose the metric perturbations into scalar modes, vector modes, and tensor modes. The helicities completely decouple at linear perturbation level.  We define the metric perturbations as follows,
\begin{eqnarray}
g_{00}&=&-\left(1+2\phi\right)~,\nonumber\\
g_{0i}&=&a(t)\left(S_i+\partial_i\beta\right)~,\nonumber\\
g_{ij}&=&a^2(t)\left[\delta_{ij}+2\psi\delta_{ij}+(\partial_i\partial_j-\frac{1}{3}\partial^2)E+\frac{1}{2}(\partial_iF_j+\partial_jF_i)+\gamma_{ij}\right]~,
\end{eqnarray}
where $\phi,~\beta,~\psi,~E$ are scalar perturbations, $S_i,~F_i$ are vector perturbations and $\gamma_{ij}$ is the tensor perturbations. Vector modes satisfy the transverse condition,
\begin{eqnarray}
\partial_iS^i=\partial_iF^i=0~.
\end{eqnarray}
Tensor modes satisfy transverse and traceless condition, 
\begin{eqnarray}
\gamma^i_i=\partial_i\gamma^{ij}=0~.
\end{eqnarray}
Under spatial diffeomorphisms, the vector field defined by \cite{Gumrukcuoglu:2011zh}
\begin{eqnarray}
Z^i\equiv\frac{1}{2}\delta^{ij}(\partial_j E+F_j)
\end{eqnarray}
transforms as 
\begin{eqnarray}
Z^i\to Z^i+\xi^i~.
\end{eqnarray}
Comparing to Eq. (\ref{pidif}), we see that the combination $(Z^i+\pi^i)$ is a gauge invariant quantity. In the unitary gauge, $Z^i$ eats $\pi^i$, and survives in the linear perturbation theory. This is  in contrast to general relativity, where both of $E$ and $F_i$ are non-dynamical and can be integrated out. 

In this section, we analyse the metric perturbations in unitary gauge, in which we fix $\pi^i=0$. 

\subsubsection{Scalar Modes}
In the scalar sector, $\phi$, $\beta$ and $\psi$ are non-dynamical. After integrating them out, we obtain the quadratic action for scalar modes,
\begin{eqnarray}
S_{s}^{(2)}=M_p^2\int dtd^3k\:\left(\mathcal{K}_s\dot{E}^2-\Omega_sE^2\right)~,
\end{eqnarray}
where 
\begin{eqnarray}\label{scalardis}
\mathcal{K}_s&=&\frac{-k^4a^5\dot{H}}{4\left(k^2-3a^2\dot{H}\right)},\nonumber\\
\Omega_s&=&\frac{a^3 k^4  \left[k^4 \ddot{H}-H \dot{H} \left(36 a^4 \dot{H}^2-21 a^2 k^2 \dot{H}+k^4\right)\right]}{12 H \left(k^2-3 a^2 \dot{H}\right)^2}+\frac{2k^4M_2^2}{3a}.
\end{eqnarray}
The scalar mode is ghost free, as long as $\dot{H}$ is negative. On the other hand, the kinetic term vanishes in the limit $\dot{H}\to0$, which implies the strong coupling in this background, as found in the previous subsection. 
The scalar action is canonically normalized by defining the field as 
\begin{eqnarray}
\mathcal{E}\equiv \left(\frac{-M_p^2k^4a^2\dot{H}}{2k^2-6a^2\dot{H}}\right)^{1/2}E,
\end{eqnarray}
with the result that
\begin{eqnarray}
S_{s}^{(2)}=\frac{1}{2}\int dtd^3ka^3\left(\dot{\mathcal{E}}^2-\omega_s^2\mathcal{E}^2\right),
\end{eqnarray}
where 
\begin{eqnarray}\label{sdis}
\omega_s^2=&&\frac{8M_2^2}{a^4}+\frac{8 k^2M_2^2}{3 a^6 H^2 \epsilon }+\frac{36 a^4 H^6 \epsilon ^3}{\left(k^2+3 a^2 H^2 \epsilon \right)^2}+\frac{3 a^2 H^4 k^2 \epsilon  \left(\eta ^2+\eta -\eta  s+2 \epsilon ^2-(\eta -22) \epsilon -2\right)}{2 \left(k^2+3 a^2 H^2 \epsilon \right)^2}\nonumber\\
&&-\frac{H^2 k^4 \left(\eta ^2+2 \eta  (s+5)-(6 \eta +56) \epsilon +16\right)}{4 \left(k^2+3 a^2 H^2 \epsilon \right)^2}+\frac{k^6 (1 +2 \epsilon -\eta)}{3 a^2 \left(k^2+3 a^2 H^2 \epsilon \right)^2}~,
\end{eqnarray}
The ``slow roll" parameters used here are defined by 
\begin{eqnarray}
\epsilon\equiv-\frac{\dot{H}}{H^2},~~~~~~
\eta\equiv\frac{\dot{\epsilon}}{H\epsilon},~~~~~~
s\equiv\frac{\dot{\eta}}{H\eta}.
\end{eqnarray}
In the IR regime  $k^2\ll a^2H^2\epsilon$, the dispersion relation of scalar modes can be perturbatively expanded with respect to $k$, revealing a relativistic dispersion relation
\begin{eqnarray}
\omega_s^2\simeq \frac{c_s^2k^2}{a^2}+m_s^2,
\end{eqnarray}
where
\begin{eqnarray}\label{IRscalar}
c_s^2&\equiv&1+\frac{\epsilon}{3}-\frac{\eta}{6}-\frac{2-\eta^2-\eta+\eta s}{6\epsilon}+\frac{8M_2^2}{3a^4H^2\epsilon},\nonumber\\
m_s^2&\equiv&4 H^2 \epsilon+\frac{8 M_2^2}{a^4}~.
\end{eqnarray}
though the speed of sound at momenta $k>m_s$ differs from the speed of light.  The mass and sound speed receive both model independent and model dependent contributions.   In general $M_2^2$ is a function of $g^{ii}$ and thus scale factor and time dependent. If $M_2^2\propto a^4$, the mass of scalar mode will approach to a constant value at late times. 

At short distance, $k^2\gg a^2H^2$, the dispersion relation simplifies to
\begin{eqnarray}\label{scaldispersionhik}
\omega_s^2\simeq\frac{k^2}{a^2}\cdot \left(\frac{8 M_2^2}{3 a^4 H^2 \epsilon }+\frac{ 1+2 \epsilon-\eta}{3 }\right)+\frac{8 M_2^2}{a^4}-\frac{1}{4} H^2 \left[\eta ^2+2 \eta  (s+5)+16 \epsilon ^2-2 (7 \eta +24) \epsilon +16\right].
\end{eqnarray}
At short wavelength, the leading order of mass term in the dispersion relation can also be neglected. After doing so, this result agrees with the one calculated in the Goldstone gauge, i.e. eq. (\ref{decoupling}) under the helicity decomposition $\pi^i=\partial^i\varphi+A^i$ where $\varphi$ is the scalar mode considered just now in the unitary gauge.

\subsubsection{Vector Modes}
In the vector sector, $S_i$ is non-dynamical. After integrating it out, the effective action for the vector degrees of freedom reads
\begin{eqnarray}
S_v^{(2)}=M_p^2\int \mathcal{K}_v\dot{F}_i\dot{F}^i-\Omega_vF_iF^i,
\end{eqnarray}
where
\begin{eqnarray}\label{vectordis}
\mathcal{K}_v&=&\frac{-k^2 a^5  \dot{H}}{4 \left(k^2-4 a^2\dot{H}\right)},\nonumber\\
\Omega_v&=&-\frac{1}{4}k^2a^3\dot{H}+\frac{M_2^2k^2}{2a}.
\end{eqnarray}
Similar to the scalar modes, the vector modes are free from ghost instability when $\dot{H}<0$.  We define the canonical field variable, 
\begin{eqnarray}
\mathcal{F}_i\equiv \left(\frac{-M_p^2k^2a^2\dot{H}}{2k^2-8a^2\dot{H}}\right)^{1/2}F_i~.
\end{eqnarray}
and the canonically normalized action is
\begin{eqnarray}
S_v^{(2)}=\frac{1}{2}\int dtd^3ka^3\left(\dot{\mathcal{F}}_i\dot{\mathcal{F}}^i-\omega_v^2\mathcal{F}_i\mathcal{F}^i\right)
\end{eqnarray}
where 
\begin{eqnarray}
\omega_v^2=&&\frac{8 M_2^2}{a^4}+\frac{2 k^2 M_2^2}{a^6 H^2 \epsilon }+\frac{64 a^4 H^6 \epsilon ^3}{ \left(4 a^2 H^2 \epsilon +k^2\right)^2}+\frac{2 a^2 H^4 k^2 \epsilon  \left(\eta ^2+\eta -\eta  s+2 \epsilon ^2-(\eta -24) \epsilon -2\right)}{\left(4 a^2 H^2 \epsilon +k^2\right)^2}\nonumber\\
&&-\frac{H^2 k^4 \left(\eta ^2+2 \eta  (s+5)+8 \epsilon ^2-2 (5 \eta +36) \epsilon +16\right)}{4\left(4 a^2 H^2 \epsilon +k^2\right)^2}+\frac{k^6}{a^2 \left(4 a^2 H^2 \epsilon +k^2\right)^2}~.
\end{eqnarray}
In the IR regime $k^2\ll a^2H^2\epsilon$, the dispersion relation of vector modes is also approximated by the relativistic form
\begin{eqnarray}
\omega_v^2\simeq \frac{c_v^2k^2}{a^2}+m_v^2,
\end{eqnarray}
where
\begin{eqnarray}\label{IRvector}
c_v^2&\equiv&1+\frac{\epsilon}{4}-\frac{\eta}{8}-\frac{2-\eta-\eta^2+\eta s}{8\epsilon}+\frac{2M_2^2}{a^4H^2\epsilon},\nonumber\\
m_v^2&\equiv&\frac{8 M_2^2}{a^4}+4 H^2 \epsilon.
\end{eqnarray}
Comparing to Eq. (\ref{IRscalar}), we see that in the IR limit, the mass of vector modes is the same as that of scalar modes. 

At short distance $k^2\gg a^2H^2$, the dispersion relation of vector modes is simplified as 
\begin{eqnarray}\label{vecdispersionhik}
\omega_v\simeq \frac{k^2}{a^2} \left(1+\frac{2 M_2^2}{a^4 H^2 \epsilon }\right)+\frac{8 M_2^2}{a^4}-\frac{1}{4} H^2 \left[\eta ^2+2 \eta  (s+5)+8 \epsilon ^2-10 (\eta +4) \epsilon +16\right].
\end{eqnarray}
Note that sound speeds of scalar modes and vector modes are not independent. In the short distance $k^2\gg a^2H^2$, they are related by $c_v^2\simeq \frac{3}{4}\left(1+c_s^2\right)$, under the approximation that all slow roll parameters are much smaller than unity,  and in this limit agrees with the result of \cite{Endlich:2012pz}. This is due to the uniqueness of the quadratic operator $\bar{\delta} g^{ij}\bar{\delta} g^{ij}$ at linear perturbation level.

\subsubsection{Tensor Modes} The quadratic action for tensor modes reads 
\begin{eqnarray}\label{tensor}
S_T^{(2)}=\frac{M_p^2}{8}\int a^3\left[\dot{\gamma}_{ij}\dot{\gamma}^{ij}-\left(\frac{k^2}{a^2}+\frac{8M_2^2}{a^4}+4H^2\epsilon\right)\gamma_{ij}\gamma^{ij}\right].
\end{eqnarray}
The tensor modes also become massive, with the same mass as the scalar and vector modes in the IR regime.  In the observational aspect, the non-vanishing mass gap leads to a sharp peak on the stochastic gravitational waves spectrum. The position and height of the peak carry information on the present value of the mass term, as well as the duration of the inflationary stage \cite{Gumrukcuoglu:2012wt}.

\subsection{Higher Order Derivatives}\label{sechd}
In this subsection, we expand up to second order derivatives.  In general, we may expect higher-derivative operators to arise at least after calculating loop corrections to the action~\cite{deRham:2013qqa}. To this order, a (over)complete set of operators is
\begin{eqnarray}\label{hdterms}
&&A_1\cdot\mathcal{R},~~~
A_2\cdot\mathcal{R}^{ii},~~~
A_3\cdot\mathcal{R}^{ijij},~~~
A_4\cdot\nabla^{\mu}\nabla^{\nu}x^a\nabla_{\mu}\nabla_{\nu}x^b\delta_{ab}, ~~~
A_5\cdot\square x^a\square x^b\delta_{ab},~~~
A_6\cdot\nabla^{\mu}A_7\nabla_{\mu}A_8,
\nonumber\\
&&
A_9f_{\mu\nu}\nabla^\mu A_{10}\nabla^\nu A_{11},~~~
A_{10}\cdot\nabla^{\mu}\nabla^{\nu}x^a\nabla^{\rho}\nabla^{\sigma}x^b\delta_{ab}\cdot f_{\mu\nu}f_{\rho\sigma},~~~
A_{11}\cdot\nabla^{\mu}\nabla^{\nu}x^a\nabla^{\rho}\nabla^{\sigma}x^b\delta_{ab}\cdot f_{\mu\rho}f_{\nu\sigma},
\nonumber\\
&&A_{12}f_{\mu\nu}(\nabla^{\mu}f_{\rho\sigma})(\nabla^\nu x^a)\nabla^\rho\nabla^\sigma x^b\delta_{ab},~~~
A_{13}(\nabla^{\mu}f_{\rho\sigma})(\nabla^\rho f_{\mu\nu})(\nabla^\nu x^a)(\nabla^\sigma x^b)\delta_{ab}, ~~~\partial_{\mu}\bar{\delta}g^{ij}\partial^{\mu}\bar{\delta}g^{ij},
\end{eqnarray}
where $A_n$s are generic functions of $g^{\mu\nu}$ and $f_{\mu\nu}\equiv\partial_{\mu}x^a\partial_{\nu}x^b\delta_{ab}$.  Compared to first order derivative terms, these terms are suppressed by the UV scale $\Lambda^{-2}$ and thus less relevant at low energy.  

By means of the following metric field redefinition, a generalised conformal transformation,
\begin{eqnarray}
g_{\mu\nu}\to\left(1+B\right)g_{\mu\nu}\equiv \tilde g_{\mu\nu},
\end{eqnarray}
a theory with second order derivatives is mapped to a theory with first order derivative terms plus third and higher derivatives 
\begin{eqnarray}
\int d^4x\sqrt{-g}\left[\frac{1}{2}M_p^2\mathcal{R}+F\left(g^{\mu\nu},f_{\mu\nu}\right)+G\left(\nabla^{\mu}\nabla^{\nu}x^a,...\right)\right]\to
\int d^4x\sqrt{-\tilde g}\left[\frac{1}{2}M_p^2\mathcal{R}+\tilde F\left(\tilde g^{\mu\nu},f_{\mu\nu}\right)+\mathcal{O}(\nabla^3x)\right]\nonumber\\
\end{eqnarray}
where $B$ and $G$ are the functions of higher order derivative terms and 
\begin{eqnarray}
B=A\left(g^{\mu\nu},f_{\mu\nu}\right)\cdot G,
\end{eqnarray}
where $A\left(g^{\mu\nu},f_{\mu\nu}\right)$ is a function of $g^{\mu\nu}~\text{and}~f_{\mu\nu}$, and its form is decided by the parameters in $F\left(g^{\mu\nu},f_{\mu\nu}\right)$. This field redefinition works equally well if the next leading derivative terms are third-order, with the resulting theory containing only first-order derivatives and fourth- and higher-order derivatives.  The procedure could be repeated to remove derivatives up to a desired finite order: Starting at $n\geq 2$, $n$-order derivative operators can be removed in favor of $n+1$-order derivative and higher terms.

As an example, consider an action with two second-derivative terms,
\begin{eqnarray}
\int\sqrt{-g}\left[\frac{1}{2}M_p^2\mathcal{R}+M_p^2m^2\left(c_0+c_1f+c_2f^2+c_3f^3\right)+M_pm\left(\nabla^{\mu}\nabla^{\nu}x^a\nabla_{\mu}\nabla_{\nu}x^b\delta_{ab}-\square x^a\square x^b\delta_{ab}\right)\right]\nonumber\\
\end{eqnarray}
where $f\equiv g^{\mu\nu}f_{\mu\nu}$.  It is equivalent to 
\begin{eqnarray}
\int\sqrt{-\tilde{g}}\left[\frac{1}{2}M_p^2\tilde{\mathcal{R}}+M_p^2m^2\left(c_0+c_1\tilde{f}+c_2\tilde{f}^2+c_3\tilde{f}^3\right)\right]
\end{eqnarray}
with 
\begin{eqnarray}\label{gtilde}
\tilde{g}_{\mu\nu}\equiv \left[1+ \frac{\left(\nabla^{\mu}\nabla^{\nu}x^a\nabla_{\mu}\nabla_{\nu}x^b\delta_{ab}-\square x^a\square x^b\delta_{ab}\right)}{M_pm\left(2c_0+c_1f-c_3f^3\right)}\right]g_{\mu\nu}.
\end{eqnarray}
We have used the approximation 
\begin{eqnarray}
\sqrt{-g}\left[1+ \frac{\left(\nabla^{\mu}\nabla^{\nu}x^a\nabla_{\mu}\nabla_{\nu}x^b\delta_{ab}-\square x^a\square x^b\delta_{ab}\right)}{M_pm\left(2c_0+c_1f-c_3f^3\right)}\right]\mathcal{R}\simeq \sqrt{-g}\mathcal{R},
\end{eqnarray}
because terms like $\square x^a\square x^b\delta_{ab}\cdot\mathcal{R}$ are third order derivative terms, so that they are additionally suppressed and we can neglect these terms when truncating at the second order derivatives. The effective action for Goldstone bosons is derived from eq. (\ref{ggact}) with $H^2$, $\dot{H}$ and $g^{ij}$ replaced by the ones induced by Eq. (\ref{gtilde}), and then performing the spatial diffeomorphism transformation shown in Sec.\,\ref{Goldstone}.

Notably this field redefinition implies that the sound speeds of the scalar and vector modes (at $k^2\gg a^2H^2$, Eqs.\,\eqref{scaldispersionhik} and \eqref{vecdispersionhik} respectively) are modified only by small corrections to the cosmological parameters $H,\epsilon,\eta,s$ due to the change in metric.  Since the expressions Eqs.\,\eqref{scaldispersionhik} and \eqref{vecdispersionhik} remain valid, the high energy relation 
\beqn
c_v^2\simeq \frac{3}{4}\left(1+c_s^2\right)
\eeqn 
is preserved even in the presence of second-order derivative terms.  Repeating the metric redefinition procedure to remove derivatives terms of any finite order, we find that the relation is a robust prediction of the effective theory, valid as long as the underlying derivative expansion is valid.

\subsection{de Sitter and Minkowski limits}\label{sec:dSMink}

The higher derivative terms become important in the de Sitter $\dot{H}\to0$ and Minkowskian $\dot{H}=H^2\to 0$ limit. In the limit, the kinetic term from the lowest dimensional operators vanishes, and kinetic terms arising from higher derivative operators become leading order.  This eliminates the strong coupling problem in the de Sitter and Minkowski limits.

For instance, with higher order derivative term $\partial_{\mu}\bar{\delta}g^{ij}\partial^{\mu}\bar{\delta}g^{ij}$ taken into account, in Minkowskian limit $\dot{H}=H^2\to0$, we have 
\begin{eqnarray}
S&=&\int d^4x\sqrt{-g}\left[-M_p^2M_2^2\bar{\delta}g^{ij}\bar{\delta}g^{ij}-d_2 M_pM_2\partial_{\mu}\bar{\delta}g^{ij}\partial^{\mu}\bar{\delta}g^{ij}+...\right]\nonumber\\
&=&\int d^4x\sqrt{-g}\left[2d_2 M_pM_2\left(\partial_i\dot{\pi}^j\partial_i\dot{\pi}^j+\frac{1}{3}\left(\partial_i\dot{\pi}^i\right)^2-\partial^2\pi^i\partial^2\pi^i-\frac{1}{3}\partial_i\partial_j\pi^j\partial_i\partial_k\pi^k\right)\right.
\nonumber\\ \label{GSactionHD}
&&\hspace{60mm}\left.-2M_p^2M_2^2\cdot\left(\partial_i\pi^a\partial_i\pi^a+\frac{1}{3}\left(\partial_a\pi^a\right)^2\right)+...\right],
\end{eqnarray}
where $\partial^2\equiv \partial_i\partial_j\delta_{ij}$.  $d_2$ is an $\mathcal{O}(1)$ positive constant since a priori higher derivative terms may be similar in size to the leading term and suppressed primarily by the additional powers of $k^2/\Lambda^2$.  The Goldstone action shows clearly how the goldstones obtain non-vanishing kinetic terms directly related to the higher derivative terms.  
After canonical normalization, we see that instead of strong coupling, the corresponding scalar and vector modes become massive in the Minkowskian space-time, with masses $m^2\sim d_2M_pM_2$.  

For the tensor modes, the situation is different: the quadratic term $M_2^2\bar{\delta}g^{ij}\bar{\delta}g^{ij}$ still provides a nonvanishing mass, as seen in the $H\to 0$ limit of Eq.\,\eqref{tensor}.  The higher derivative terms only make a correction to the mass.  
For instance, if we include the higher order term  $\partial_{\mu}\bar{\delta}g^{ij}\partial^{\mu}\bar{\delta}g^{ij}$ in the Minkowskian space-time limit, the tensor action is
\begin{eqnarray}
S&\supset&\frac{M_p^2}{8}\int\left(1+\frac{8d_2M_2}{M_p}\right)\left(\dot{\gamma}_{ij}\dot{\gamma}^{ij}-k^2\gamma_{ij}\gamma^{ij}\right)-8M_2^2\gamma_{ij}\gamma^{ij},
\end{eqnarray}
Canonically normalizing $\gamma^{ij}$, the graviton mass is
\begin{eqnarray}
m_T^2\simeq 8M_2^2(1-\frac{8d_2M_2}{M_p}).
\end{eqnarray}
Consequently, the vector and scalar modes have masses $m_v,m_s\sim\sqrt{M_p/M_2}\cdot m_T$.  Considering horizon-scale perturbations $k\sim H$, these modes are relatively heavy and could be integrated out, having $\sim1/M_p^2$ impact on tensor-mode observables.  On the other hand, we would like to informatively mention that the sound speed of tensor mode will also be modified, with more of higher derivative terms in eq. (\ref{hdterms}) included. In this case, the sound speed of tensor mode deviates from unity by a factor of $M_2/M_p$.
\\
\\
To conclude, in this section we have investigated perturbations on the FRW background. We first derived the Goldstone action up to quadratic order, which clearly isolated the strong coupling problem as well as a possible resolution by the inclusion of higher derivative terms.  Seeing that mixing with the metric can not in general be neglected, we then performed a full perturbation analysis in the unitary gauge. This analysis exhibited a well-behaved massive spin two particle, with 5 polarizations: one scalar mode, two vector modes, and two tensor modes. All helicity modes are massive, and the masses shown to be identical in the low momentum regime.  The dispersion relations of these 5 modes are fully characterized by the parameter set $\{H, \epsilon, \eta, s, M_2^2\}$.

\section{Several Examples}
 
\subsection{The Minimal Model and Next-to-minimal Model}\label{Eg:mini}
The simplest example is obtained by setting $M_2^2=0$ in the general action Eq. (\ref{ggact}), and only keeping the first three terms. In the $\phi^a$ language, this theory corresponds to Einstein gravity and three canonical massless scalar fields with space like VEV \req{scond} \cite{Lin:2013sja},
\begin{eqnarray}\label{minimal}
S=M_p^2\int d^4 x \sqrt{-g}\left(\frac{\mathcal{R}}{2}-\frac{1}{2}m^2g^{\mu\nu}\partial_{\mu}\phi^a\partial_{\nu}\phi^b\delta_{ab}-\Lambda\right),
\end{eqnarray}
where $\Lambda$ is the bare cosmological constant.
The energy density of the spatial condensate scales as $\rho\propto a^{-2}$ and its equation of state equals to $-1/3$.  In the linear perturbation theory, after canonical normalization, scalar, vector, and tensor polarizations of graviton have the same dispersion relations
\begin{eqnarray}\label{sdisp}
\omega_s^2=\omega_v^2=\omega_t^2= \frac{k^2}{a^2}+\frac{2m^2}{a^2}~,
\end{eqnarray}
with the same non-vanishing mass. These dispersion relations are identical due to the $SO(3)$ internal symmetry of the scalar fields, which has been imposed to ensure the rotational symmetry of the vev configuration. For the same reason, the scalars can be re-decomposed into 3 polarizations: two transverse modes and one longitudinal mode,
\begin{eqnarray}
\pi^a=\delta^{a i}\left(\partial_{i}\varphi+A_{i}\right),
\end{eqnarray}
where $\partial_iA^i=0$. Due to the $SO(3)$ symmetry, we could rotate longitudinal mode a bit ``into" transverse modes, and on the other hand transverse modes are rotated a bit ``into" longitudinal mode, and leave the action invariant. In the unitary gauge, these transverse and longitudinal modes are eaten by graviton, It implies the masses of scalar modes and vector modes of  graviton should be the same. 

To see how the effective theory operator $\bar\delta g^{ij}\bar\delta g^{ij}$ in \req{GSaction} is related to a specific model, we consider as an example a general polynomial of the tensor defined in Eq. (\ref{fidtensor}), i.e. $f_{\mu\nu}\equiv\partial_{\mu}x^a\partial_{\nu}x^b\delta_{ab}, ~f\equiv g^{\mu\nu}f_{\mu\nu}$. For instance, starting from the  theory truncated at cubic order,
\begin{eqnarray}
S=M_p^2\int d^4x \sqrt{-g}\left[\frac{\mathcal{R}}{2}-m^2\left(c_0+c_1f+c_2f^2+d_2f^{\mu}_{~\nu}f^{\nu}_{\mu~}+c_3f^3+d_3f^{\mu}_{~\rho}f^{\rho}_{~\sigma}f^{\sigma}_{~\mu}+g_3f\cdot f^{\mu}_{~\nu}f^{\nu}_{\mu~}\right)\right],
\end{eqnarray}
where $c_0$ is the bare cosmological constant. In this case, the coefficient $M_2^2$ equals to 
\begin{eqnarray}
M_2^2=m^2\left(d_2+\frac{3 d_3 }{a^2}+\frac{3 g_3 }{a^2}\right).
\end{eqnarray}
The dispersion relations of the 5 polarizations of graviton can then be read from Eq. (\ref{scalardis})(\ref{IRscalar}), Eq. (\ref{vectordis})(\ref{IRvector}) and Eq. (\ref{tensor}) directly. 

Generally, given a Lagrangian with the function of $-\int\sqrt{-g}F\left(g^{\mu\nu},f_{\mu\nu}\right)$, we can calculate $M_2^2$ in this way: we first Taylor expand the Lagrangian around background, 
\begin{eqnarray}
F\left(g^{\mu\nu},f_{\mu\nu}\right)=F_0+\frac{\delta F}{\delta g^{\mu\nu}}\delta g^{\mu\nu} +\frac{1}{2!} \frac{\delta^2F}{\delta g^{\mu\nu}\delta g^{\rho\sigma}}\delta g^{\mu\nu}\delta g^{\rho\sigma}+....
\end{eqnarray}
Then note that at linear perturbation level, $\bar{\delta}g^{ij}$ equals the trace-less part of metric fluctuation $\delta g^{ij}$. Finally, we decompose the metric fluctuation into trace part and trace-less part, $\delta g^{ij}=\bar{\delta}g^{ij}+\frac{1}{3}\delta g^{kk}\delta^{ij}$. $M_2^2$ is identified as the coefficient in front of the sum of the trace-less terms (see the appendix \ref{app:M2} for more details).

\subsection{Generalization to Spatially Non-flat Universe}
Up to now, we have analysed the gravity theory with broken spatial diffeomorphisms in a flat FRW universe. It is straightforward to generalize it to a non-flat FRW universe. In this case, the internal metric of scalar fields configuration is replaced by one which is compatible with the metric on the non-flat spatial slice.
 
For a spatially non-flat FRW universe, the space time metric can be written as 
\begin{eqnarray}
ds^2=-dt^2+a(t)^2\Omega_{ij}dx^idx^j,
\end{eqnarray}
where $\Omega_{ij}dx^idx^j$ is the metric on a 3-sphere 
 \begin{eqnarray}
\Omega_{ij}\equiv \delta_{ij}+\frac{K\delta_{il}\delta_{jm}x^lx^m}{1-K\delta_{lm}x^lx^m},
\end{eqnarray}
where $K=1$ for a closed universe and $K=-1$ for an open universe. In the unitary gauge, the tensor $f_{\mu\nu}$ takes the form which compatible with 3-sphere metric,
\begin{eqnarray}
f_{\mu\nu}\equiv \partial_{\mu}\phi^a\partial_{\nu}\phi^bG_{ab}(\phi^a)=(0,\Omega_{ij}).
\end{eqnarray}
A possible vacuum configuration for scalar fields is 
\begin{eqnarray}
\phi^a=x^a,~~~~~~~G_{ab}(\phi)=\delta_{ab}+\frac{K\delta_{ac}\delta_{bd}\phi^c\phi^d}{1-K\delta_{cd}\phi^c\phi^d}.
\end{eqnarray}
It is easy to check that the above vacuum configuration are indeed on shell and satisfy the equations of motion,
\begin{eqnarray}
g^{\mu\nu}\nabla_{\mu}\nabla_{\nu}\phi^a+g^{\mu\nu}\partial_{\mu}\phi^b\partial_{\nu}\phi^c\Gamma^a_{bc}=0,
\end{eqnarray}
where $\Gamma^a_{bc}$ is the affine connection which derived from the inner metric $G_{ab}(\phi)$. The generalization of our effective field theory approach to a spatially non-flat universe is quite straight forward. Including spatial curvature, the effective action can be written as 
\begin{eqnarray}
S=M_p^2\int d^4x\sqrt{-g}\left[\frac{1}{2}\mathcal{R}-3\left(H^2+\dot{H}\right)+\left(a^2\dot{H}-K\right)\Omega_{ij}g^{ij}-M_2^2\bar{\delta} g^{ij}\bar{\delta} g^{ij}+...\right],
\end{eqnarray}
where the quadratic order operator is defined by 
\begin{eqnarray}
\bar{\delta} g^{ij}\equiv g^{ik}\Omega_{kj}-3\frac{g^{ik}g^{lm}\Omega_{kl}\Omega_{mj}}{g^{ij}\Omega_{ij}}.
\end{eqnarray}
Non-zero spatial curvature is sufficient to ensure the kinetic term is non-degenerate.  This suggests that another way to cure the strong coupling problem of massive gravity in Minkowski space is to perturb in the direction of non-vanishing spatial curvature.

\subsection{A Self-accelerating Universe}
When we apply our massive gravity theory to cosmology, one of most interesting questions is whether or not a graviton mass term can accelerate the cosmic expansion.  A similar question was studied in Ref. \cite{Endlich:2012pz}, in which they proposed an inflationary model ``solid inflation'', with de Sitter-like expansion driven by the vacuum energy of the ``solid", that is the spatial condensate vacuum configuration \req{scond}. In this section, we provide another way to realize a de Sitter phase. 

We work on the static chart of the de Sitter phase, where the metric takes the form 
\begin{eqnarray}\label{staticm}
ds^2=-(1-H^2r^2)dt^2+\frac{1}{1-H^2r^2}dr^2+r^2d\theta^2+r^2\sin\theta^2d\phi^2~,
\end{eqnarray}
and $H$ is the Hubble constant of de-sitter space-time. The Einstein tensor reads,
\begin{eqnarray}\label{staticET}
G^{\mu}_{\nu}=-3H^2\delta^{\mu}_{\nu}~.
\end{eqnarray}
In terms of spherical coordinate, the 3 scalars can be written as 
\begin{eqnarray}\label{qsc}
\phi^a=x^a~,~~~~x^a=r(\sin \theta  \cos \phi ,\sin \theta  \sin \phi ,\cos \theta )~,
\end{eqnarray}
and the tensor 
\begin{eqnarray}
f_{\mu\nu}\equiv\partial_{\mu}\phi^a\partial_{\nu}\phi^b\delta_{ab}=(0,1,r^2,r^2\sin\theta^2)~.
\end{eqnarray}
Normally, such a field configuration is not compatible with the space time metric Eq. (\ref{staticm}), since the scalars do not satisfy their equation of motion. For instance, if we consider a canonical Lagrangian,
\begin{eqnarray}
 M_p^2\int d^4x\sqrt{-g}\left\{\frac{\mathcal{R}}{2}-\frac{1}{2}m^2g^{\mu\nu}\partial_{\mu}\phi^a\partial_{\nu}\phi^b\delta_{ab}\right\}~,
\end{eqnarray}
the energy momentum tensor reads, 
\begin{eqnarray}
T^{0}_{~0}&=&\frac{1}{2} H^2 r^2-\frac{3}{2},\nonumber\\
T^{1}_{~1}&=&-\frac{1}{2} H^2 r^2-\frac{1}{2},\nonumber\\
T^{2}_{~2}&=&\frac{1}{2}H^2 r^2-\frac{1}{2},\nonumber\\
T^{3}_{~3}&=&T^{2}_{~2}.
\end{eqnarray}
Comparing to Eq.(\ref{staticET}), we can see Einstein equations are not satisfied. On the other hand, the vacuum configuration does not satisfy the equations of motion for scalar fields  $\phi^a$ either,
\begin{eqnarray}
g^{\mu\nu}\nabla_{\mu}\nabla{_\nu}\phi^a\propto H^2r \neq 0.
\end{eqnarray}
The $SO(3)$ scalar fields' configuration in the static chart of de Sitter phase implies that there are large shears and energy momentum flows in the non-static coordinates.  Nevertheless, if we include the higher order kinetic interaction terms, by tuning the parameters  we may be able to cancel out the shears and flows and realize a  self-consistent de Sitter solution.  We found this solution at least contains  4th order kinetic interactions,
\begin{eqnarray}
M_p^2m^2\int d^4x\sqrt{-g}&&\left[c_1f+c_{21}f^2+c_{22}f^{\mu}_{\nu}f^{\nu}_{\mu}+c_{31}f^3+c_{32}f^{\mu}_{\nu}f^{\nu}_{\rho}f^{\rho}_{\mu}+c_{33}f\cdot f^{\mu}_{\nu}f^{\nu}_{\mu}\right.\nonumber\\
&&c_{41}f^4+c_{42}f^{\mu}_{\nu}f^{\nu}_{\rho}f^{\rho}_{\lambda}f^{\lambda}_{\mu}+c_{43}f\cdot f^{\mu}_{\nu}f^{\nu}_{\rho}f^{\rho}_{\mu}+c_{44}(f^{\mu}_{\nu}f^{\nu}_{\mu})^2+c_{45}f^2\cdot f^{\mu}_{\nu}f^{\nu}_{\mu}\left.\right],
\end{eqnarray}
where all coefficients $c_1,~c_{21},~c_{22}...$ are coordinate independent constants. As a self-consistent solution, it must satisfy Einstein equations, $T^{\mu}_{\nu}=G^{\mu}_{\nu}=-3H^2\delta^{\mu}_{\nu}$,  and the equation of motion for scalar fields as well. These conditions lead to a set of nontrivial constraints on the coefficients,
\begin{eqnarray}\label{coefficients}
c_{21}&=& \frac{1}{8} \left(-3 c_1-2 c_{33}\right),\nonumber\\
c_{22}&=& \frac{3}{8} \left(c_1-2 c_{33}\right),\nonumber\\
c_{41}&=& \frac{1}{48} c_1-\frac{1}{24}  \left(c_{33}+4 c_{45}\right),\nonumber\\
c_{42}&=& \frac{3 c_1}{16}+c_{45}-\frac{c_{33}}{8},\nonumber\\
c_{43}&=& \frac{1}{6}  \left(c_{33}-8 c_{45}\right)-\frac{1}{12} c_1,~\nonumber\\
c_{44}&=& \frac{1}{8} \left(-c_1-4 c_{45}\right),\nonumber\\
c_{32}&=& -c_{33},\nonumber\\
c_{31}&=&0~,
\end{eqnarray}
and the energy momentum tensor can be calculated as 
\begin{eqnarray}
T^{\mu}_{\nu}=\frac{3}{8} \left(3 c_1-2 c_{33}\right)M_p^2m^2\delta^{\mu}_{\nu}~.
\end{eqnarray}
Indeed with this choice of parameters, the vacuum energy behaves as an effective cosmological constant.

However, as we mentioned in the section III. B, in the de Sitter limit, Goldstone bosons become strongly coupled, and the theory breaks down. To show how it occurs in this model, we perturb the scalar fields' configuration by introducing the (Goldstone) excitations 
\begin{eqnarray}
\phi^a=x^a+\pi^a~.
\end{eqnarray}
Then we take the decoupling limit to decouple Goldstone bosons and gravity,
\begin{eqnarray}
m\to0, ~~~M_p\to\infty,~~~\Lambda_2\equiv M_pm=\,\mathrm{const}.
\end{eqnarray}
In momentum space, the quadratic action for Goldstone bosons reads
\begin{eqnarray}
\mathcal{L}^{(2)}_{\pi}=\int \mathcal{K}\delta_{ab}\dot{\pi}^a\dot{\pi}^b-c_{diag}^2k^2\pi^a\pi^b\delta_{ab}-c_{mix}^2(k_a\pi^a)^2~,
\end{eqnarray}
where 
\begin{eqnarray}
\mathcal{K}&=&-\left(c_1+6 c_{21}+2 c_{22}+27 c_{31}+3 c_{32}+9 c_{33}+108 c_{41}+4 c_{42}+12 c_{43}+12 c_{44}+36 c_{45}\right)~,\nonumber\\
c_{diag}^2&=&-\left(c_1+6 c_{21}+4 c_{22}+27 c_{31}+9 c_{32}+15 c_{33}+108 c_{41}+16 c_{42}+30 c_{43}+24 c_{44}+54 c_{45}\right)~,\nonumber\\
c_{mix}^2&=&2 \left(2 c_{21}+c_{22}+18 c_{31}+3 c_{32}+7 c_{33}+108 c_{41}+6 c_{42}+15 c_{43}+14 c_{44}+39 c_{45}\right).
\end{eqnarray}
After inputting the constraints on these coefficients, i.e. Eq.\,(\ref{coefficients}), we find that $\mathcal{K}=c_{diag}^2=c_{mix}^2=0$ and the quadratic action vanishes identically. 

As we saw above in the effective theory analysis, this degeneracy arises in the exact $\dot H=0$ limit, and one of possible solutions to the strong coupling problem is to introduce a small deviation from de Sitter space time.  Another possible solution is to include higher order derivative terms. As discussed in section \ref{sec:dSMink}, the higher order derivative terms give rise to a non-vanishing kinetic term for Goldstone bosons even in de Sitter spacetime and thus heals the strong coupling problem. 

\section{Conclusion and Discussion}
In this paper, we characterize the most general theory of spatial diffeomorphisms breaking. By means of effective field theory approach,  after writing down all possible operators compatible with underlying symmetries, we are able to describe the effective theory of fluctuations around the  FRW background with spatial diffeomorphisms breaking. We showed that the most generic action on a FRW background can be written in the form 
\begin{eqnarray}
S=\int d^4x\sqrt{-g}\left[\frac{1}{2}M_p^2\mathcal{R}-3M_p^2\left(H^2+\dot{H}\right)+M_p^2a^2\dot{H}g^{ii}-M_p^2M_2^2\bar{\delta} g^{ij}\bar{\delta} g^{ij}+...\right],
\end{eqnarray}
where $\bar{\delta} g^{ij}$ is the covariant operator constructed out of $g^{ij}$ and starts from linear order in fluctuations. Differences among models (UV completions of the theory) are encoded in this operator as well as higher order operators.   The three broken spatial diffeomorphisms acquire three Goldstone bosons.  The couplings between different helicity modes are characterised by the scales $\dot{H}\text{~and~} M_2^4/\dot{H}$. 
  Above the mixing scale, we can neglect the couplings between different helicity modes, and the action of Goldstone bosons dramatically simplifies to
 \begin{eqnarray}
S=M_p^2\int d^4x\sqrt{-g}\left[a^2\dot{H}g^{\mu\nu}\partial_{\mu}\pi^a\partial_{\nu}\pi^a-\frac{1}{3}\left(2\dot{H}+\frac{\ddot{H}}{H}\right)\left(\partial_a\pi^a\right)^2-\frac{2M_2^2}{a^4}\cdot\left(\partial_i\pi^a\partial_i\pi^a+\frac{1}{3}\left(\partial_a\pi^a\right)^2\right)+...\right].\nonumber\\
\end{eqnarray}
In the unitary gauge, the 3 Goldstone bosons are eaten by graviton, and the graviton becomes a massive spin-2 particle with 5 well-behaved degrees of freedom that are decomposed into 1 scalar mode, 2 vector modes, and 2 tensor modes. 
  We performed the cosmological perturbation calculation, derived the  effective quadratic action for each of the 5 polarizations, and found that at linear perturbation level, all 5 polarisations have non-vanishing masses.  The dynamical properties of these 5 polarisations are characterised by the parameter set $\{H, \epsilon, \eta, s, M_2^2\}$, where $\epsilon, \eta, s$ are ``slow roll" parameters. 

With only first-order derivative operators, the kinetic terms of these three Goldstone bosons vanish in the de Sitter, as well as Minkowski limit, where $\dot{H}\to0$. In this limit, the kinetic terms arising from higher derivative operators become leading order. In this case, the three Goldstone bosons appear to be supermassive at the low energy scale, and thus can be integrated out, leaving two massive tensor modes as the only propagating degrees of freedom.

Away from de Sitter and Minkowski limit, in the IR regime where $k^2\ll a^2H^2\epsilon$, the quadratic action of the 5 polarisations in the unitary gauge reads
\begin{eqnarray}
&&S^{(2)}=\frac{1}{2}\int dtd^3ka^3\left[\dot{\mathcal{E}}^2-\left(\frac{c_s^2k^2}{a^2}+m_g^2\right)\mathcal{E}^2
+\dot{\mathcal{F}}_i\dot{\mathcal{F}}^i-\left(\frac{c_v^2k^2}{a^2}+m_g^2\right)\mathcal{F}_i\mathcal{F}^i+\dot{\gamma}_{ij}\dot{\gamma}^{ij}-\left(\frac{k^2}{a^2}+m_g^2\right)\gamma_{ij}\gamma^{ij}\right],\nonumber\\
\end{eqnarray}
where  $\mathcal{E}$ is the scalar mode, $\mathcal{F}_i$ is the vector mode, $\gamma^{ij}$ is the tensor mode.  The graviton mass $m_g^2=4 H^2 \epsilon+\frac{8 M_2^2}{a^4}$ takes the same value for all 5 polarisations in the long wavelength limit.  The sound speeds of scalar and vector modes are functions of ``slow roll" parameters.

Several examples on the applications of our formalism to cosmology are also presented. We started from the example of a flat FRW universe and a generic polynomial of derivative couplings and showed how to determine the dispersion relations of the 5 graviton polarizations. We also generalized to a spatially non-flat FRW universe and a static chart of de Sitter space time. 

Along this line, there are several possible extensions of our formalism. For instance, it would be necessary and more realistic to consider the matter distribution in the universe and develop a new effective theory with the coupling to mattertaken into account. On the other hand, if we go beyond linear perturbation theory, the interactions among three pions can be characterized by introducing the higher order operators like $\bar{\delta}g^{ij}\bar{\delta}g^{jk}\bar{\delta}g^{ki}$. This type of non-trivial interaction leads to non-Gaussianity in the metric fluctuations. It would be very interesting to study its possible imprint on CMB and large scale structure. 

\begin{acknowledgments}
We would like to thank R. Brandenberger, S.-H. Dai, A. De Felice, K. Izumi, X. Gao, S. Mukohyama, R. Namba, R. Saitou, M. Sasaki, A. Taruya, Y. Wang,  and  S. Weinberg for useful discussions.  Our thanks also go to the anonymous referee, for his/her useful comments and suggestions on the draft.  In this work,  C. Lin was 
 supported by Yukawa Institute for Theoretical Physics, Kyoto University, and the World Premier International Research Center Initiative (WPI Initiative), MEXT, Japan. 
\end{acknowledgments}

\appendix

\section{Calculating $M_2^2$ for a given Lagrangian}
\label{app:M2}
In the section \ref{Eg:mini}, we pointed out that for a generic Lagrangian with spatial diffeomorphisms breaking in the unitary gauge, 
\begin{eqnarray}\label{Fact}
S&\supset& -M_p^2\int\sqrt{-g} F\left(g^{\mu\nu},~\partial_{\mu}\phi^a\partial_{\nu}\phi^b\delta_{ab}\right)\nonumber\\
&=&-M_p^2\int\sqrt{-g} F\left(g^{ij}\right),
\end{eqnarray}
the associated quadratic operator in the effective field theory can be calculated by tracking the traceless part of metric perturbation. To do so, we decompose the metric as in Eq. (\ref{metricdecom}),
\begin{eqnarray}
g^{ij}\equiv a^{-2}\delta^{ij}+\frac{1}{3}\delta g^{kk}\delta^{ij}+\bar{\delta}g^{ij},
\end{eqnarray}
where $\bar{\delta}g^{ij}$ is the traceless part of metric perturbations. We expand the action Eq. (\ref{Fact}) perturbatively in terms of the above metric decomposition, 
\begin{eqnarray}\label{compare1}
S&\supset& -M_p^2\int \sqrt{-g}\left(F^{(0)}+\frac{\delta F}{\delta g^{ij}}\delta g^{ij}+\frac{1}{2!}\frac{\delta^2 F}{\delta g^{ij}\delta g^{kl}}\delta g^{ij}\delta g^{kl}+...\right)\nonumber\\
&\supset& M_p^2\int -\frac{1}{2}a^7F^{(0)}\bar{\delta} g^{ij}\bar{\delta} g^{ij}-\frac{1}{2}a^3\frac{\delta^2 F}{\delta g^{ij}\delta g^{kl}}\bar{\delta} g^{ij}\bar{\delta} g^{kl}+...\nonumber\\
&=& M_p^2\int -\frac{3}{2}a^7H^2\bar{\delta} g^{ij}\bar{\delta} g^{ij}-\frac{1}{2}a^3\frac{\delta^2 F}{\delta g^{ij}\delta g^{kl}}\bar{\delta} g^{ij}\bar{\delta} g^{kl}+...
\end{eqnarray}
where we have used the $0-0$ component of Einstein equation $3H^2=F^{(0)}$ in the last line of the above formula. 

On the other hand, the effective action for the traceless metric perturbation in terms of our effective field approach reads 
\begin{eqnarray}\label{compare2}
S&\supset&M_p^2\int\sqrt{-g}\left[-3\left(H^2+\dot{H}\right)+a^2\dot{H}g^{ii}-M_2^2\bar{\delta} g^{ij}\bar{\delta} g^{ij}+...\right]\nonumber\\
&\supset&M_p^2\int  -\frac{3}{2}a^7H^2\bar{\delta} g^{ij}\bar{\delta} g^{ij}-a^3M_2^2\bar{\delta} g^{ij}\bar{\delta} g^{ij}+...
\end{eqnarray}
By comparing eq. (\ref{compare1}) (\ref{compare2}), we find 
\begin{eqnarray}
M_2^2\bar{\delta} g^{ij}\bar{\delta} g^{ij}=\frac{1}{2}\frac{\delta^2 F}{\delta g^{ij}\delta g^{kl}}\bar{\delta} g^{ij}\bar{\delta} g^{kl}.
\end{eqnarray}

\end{document}